\documentstyle[12pt]{article}
\newcommand{\sect}[1]{\section{#1}\setcounter{equation}{0}}

\newcommand{\beq}{\begin{equation}}
\newcommand{\eeq}{\end{equation}}
\newcommand{\beqs}{\begin{eqnarray}}
\newcommand{\eeqs}{\end{eqnarray}}

\begin{document}
\begin{titlepage}
\null

\begin{flushright}
KOBE-TH-94-05\\
September 1994
\end{flushright}

\vspace{7mm}
\begin{center}
  {\Large\bf Topological Terms \par}
  {\Large\bf in String Theory on Orbifolds\footnote{Work supported
by the Grant-in-Aid for Scientific Research from the Ministry of
Education, Science and Culture (No.30183817).} \par}
  \vspace{1.5cm}
  \baselineskip=7mm

   {\large Makoto Sakamoto\footnote{E-mail address:
sakamoto@phys02.phys.kobe-u.ac.jp}\par}
\vspace{5mm}
     {\sl Department of Physics, Kobe University\\
      Rokkodai, Nada, Kobe 657, Japan \par}

\vspace{7mm}
  {\large and \par}
\vspace{7mm}

 {\large Motoi Tachibana\footnote{E-mail address:
tatibana@phys02.phys.kobe-u.ac.jp}\par}
\vspace{5mm}
  {\sl Graduate School of Science and Technology, Kobe University\\
     Rokkodai, Nada, Kobe 657, Japan \par}

\vspace{3cm}

{\large\bf Abstract}
\end{center}
\par

We study toroidal orbifold models with topologically
invariant terms in the
 path integral formalism
and give physical interpretations of the terms from
an operator formalism point of view.
We briefly discuss a possibility of a new class of
modular invariant orbifold models.

\end{titlepage}
\setcounter{footnote}{0}
\baselineskip=7mm


\sect{Introduction}

 String theory on toroidal orbifolds \cite{orb} has been
studied from both operator formalism and path  integral
formalism points of view.
Some of the advantages of the operator formalism are that the
spectrum and the algebraic structure are clear and that it is
possible to formulate the theory without Lagrangians or actions.
On the other hand, in the path integral formalism
the geometrical or topological structure is transparent and the
generalization to higher genus Riemann surfaces is obvious.
Modular invariance of partition functions is rather a trivial symmetry.
The interrelation between the two formalisms is not, however, trivial.

In ref.\cite{top}, toroidal orbifold models with topologically
nontrivial twists have been constructed in the operator formalism.
Recently, the construction in the path integral formalism
has partly
been done in ref.\cite{toroidal}. The main purpose of this paper is to
generalize the results of ref.\cite{toroidal} and to construct a wider
class of toroidal orbifold models in the path integral formalism by
adding new conformally invariant terms to the action.

A $D$-dimensional torus $T^D$ is defined by identifying
a point \{$X^I$\} with
\{$X^{I}+\pi w^I$\} for all $w^I \in \Lambda$, where $\Lambda$ is
a $D$-dimensional lattice.
An orbifold is obtained by dividing the torus by the action of a
discrete symmetry group $P$ of the torus.
Any element $g$ of $P$ can in general be represented
(for symmetric orbifolds) by \cite{orb}
\begin{equation}
g\ =\ (\ U\ ,\ v\ )\ \ ,
\end{equation}
where $U$ denotes a rotation and $v$ a shift.
In the operator formalism of closed string theory,
we can introduce a left- and right-moving coordinate
$(X^I_L,X^I_R)$.
On the orbifold, a point $(X^I_L,X^I_R)$ is identified with
$(U^{IJ}X^J_L + \pi v^I, U^{IJ}X^J_R - \pi v^I)$ for all
$(U,v) \in P.$
If we wish to formulate the orbifold model in the Polyakov
path integral formalism \cite{Polyakov}, we face two problems,
as we will explain below. Let $X^I(\sigma^1,\sigma^2)$ be a string
coordinate, which maps a Riemann surface $\Sigma$ into a target space.
In this paper, we will restrict our considerations to a genus one Riemann
surface, i.e., a torus. A generalization to higher genus Riemann surfaces
will be obvious. The string coordinate on the orbifold in general obeys
the following boundary condition:
\begin{eqnarray}
X^I(\sigma^1+1,\sigma^2) &=& U^{IJ} X^J(\sigma^1,\sigma^2) +
                             \pi w^I\ , \nonumber\\
X^I(\sigma^1,\sigma^2+1) &=&
                \widetilde{U}^{IJ} X^J(\sigma^1,\sigma^2) +
                             \pi \widetilde{w}^I\ ,
\label{twist bc}
\end{eqnarray}
for some $U,\widetilde{U} \in P$ and $w,\widetilde{w} \in \Lambda$.
The consistency of the boundary condition requires
\begin{eqnarray}
[\ U\ ,\ \widetilde{U}\ ] &=& 0\ , \\
(1-\widetilde{U})^{IJ}w^J &=& (1-U)^{IJ}\widetilde{w}^J\ .
\label{consist}
\end{eqnarray}
In the Polyakov path integral formalism, the kinetic term is given by
\begin{equation}
S_0[X,g]\ =\ \int^1_0 d^2\sigma \frac{1}{2\pi}
         \sqrt{g(\sigma)} g^{\alpha \beta}(\sigma)
         \partial_\alpha X^I(\sigma)
         \partial_\beta X^I(\sigma)\ ,
\label{action}
\end{equation}
where $g_{\alpha \beta}$ is a metric of the Riemann surface $\Sigma$
of genus one. The kinetic term is conformally invariant and is consistent
with the boundary condition (\ref{twist bc}), as it should be.
We may add the following topological term to the kinetic one:
\begin{equation}
S_{B}[X]\ =\ i\int^1_0 d^2\sigma \frac{1}{2\pi}B^{IJ}_0
\varepsilon^{\alpha \beta}\partial_\alpha X^I(\sigma)
\partial_\beta X^J(\sigma)\ ,
\label{B^{IJ}}
\end{equation}
where $\varepsilon^{\alpha \beta}$ is a totally antisymmetric tensor and
$B^{IJ}_0$ is an antisymmetric constant background field, which has been
introduced by Narain, Sarmadi and Witten \cite{N-S-W}
to explain Narain torus
compactification \cite{Narain} in the conventional approach.
The first problem is that in the path integral formalism a combination
$X^I=\frac{1}{2}(X^I_L+X^I_R)$ appears
in eqs.(\ref{action}) and (\ref{B^{IJ}}) but a combination
$\frac{1}{2}(X^I_L-X^I_R)$ does not. Hence, it seems that there is no way
to impose the twisted boundary condition corresponding to
the identification
$(X^I_L,X^I_R) \sim (U^{IJ}X^J_L+\pi v^I, U^{IJ}X^J_R-\pi v^I)$ unless
$v^I=0$ or unless we introduce a new degree of freedom corresponding to
$\frac{1}{2}(X^I_L-X^I_R)$ besides $X^I$.

The second problem is concerned with the antisymmetric background field.
The integrand of $S_{B}[X]$ is not single-valued on the Riemann surface
$\Sigma$ when the twist $U^{IJ}$ or $\widetilde{U}^{IJ}$ in
eq.(\ref{twist bc}) does not commute with $B^{IJ}_0$
\cite{I-N-T,E-J-L-M}. In ref.\cite{top}, orbifold models
with such twists have been studied in the operator formalism in detail.
The
analysis has strongly suggested that those orbifold models belong to a
topologically nontrivial class of orbifold models. However,
the topological
structure has not clearly been understood.

A solution to the above two problems has partly been given in
ref.\cite{toroidal}. In this paper, we shall propose a
more general solution applicable
to a wider class of orbifold models. Our proposal will be given in the
next section.

In section 3, we discuss physical meanings of topological
terms which we add
to the kinetic term, from a path integral formalism point of view.
The zero mode part of a one-loop partition function is computed
in the path integral formalism.

In section 4, our results in the path integral formalism are reinterpreted
from an operator formalism point of view.
We see that the interrelation between two formalism is quite
nontrivial. Section 5 is
devoted to discussion.


\sect{Topological Terms in String Theory on Orbifolds }

 In this section, we shall propose a solution to
the two problems explained
in the introduction. A key observation to solve the first problem is
a necessity of a new degree of freedom corresponding to
$\frac{1}{2}(X^I_L-X^I_R)$.
Let us introduce a new field variable $V^I(\sigma)$
which is to be regarded
as an external field. Then we have a new conformally invariant term,
\begin{equation}
S_v[X]\  =\ i\int^1_0 d^2\sigma \frac{1}{\pi} \varepsilon^{\alpha \beta}
           \partial_\alpha V^I(\sigma)\ \partial_\beta X^I(\sigma)\ .
\label{shiftaction1}
\end{equation}
Since $S_v[X]$ is independent of the metric $g_{\alpha \beta}$, it is
not only conformally invariant but also topologically invariant.
For $S_v[X]$ to be well defined on $\Sigma$, $V^I(\sigma)$ should obey
the following boundary condition:
\begin{eqnarray}
V^I(\sigma^1+1,\sigma^2) &=& U^{IJ} V^J(\sigma^1, \sigma^2)+
                             \pi v^I\ , \nonumber\\
V^I(\sigma^1,\sigma^2+1) &=&
                \widetilde{U}^{IJ} V^J(\sigma^1,\sigma^2)+
                             \pi \widetilde{v}^I\ ,
\label{shift bc}
\end{eqnarray}
for some constant vectors $v^I$ and $\widetilde{v}^I$.
The consistency of the boundary condition requires
\begin{equation}
(1-\widetilde{U})^{IJ}v^J  =  (1-U)^{IJ}\widetilde{v}^J.
\label{shift consist}
\end{equation}
Since $S_v[X]$ is a topological term, it depends only on the boundary
conditions (\ref{twist bc}) and (\ref{shift bc}). In terms of zero modes,
$S_v[X]$ can be written as
\begin{equation}
S_v[X]\ =\ -i\pi\Bigl(w^I (\widetilde{U}^T)^{IJ} \widetilde{v}^J
         -\widetilde{w}^I (U^T)^{IJ} v^J\Bigr)\ .
\label{shiftaction2}
\end{equation}
We will show in section 4 that a partition function computed in the path
integral formalism agrees with that in the operator formalism
if the vector $v^I$ in eq.(\ref{shift bc}) is identified with the shift
of the group element $g = (U,v)$.

We will next proceed to the second problem.
We shall generalize the work of
ref.\cite{toroidal}. Let $\Lambda_R(\cal G)$ ($\Lambda_W(\cal G)$) be
a root (weight) lattice of a simply-laced Lie algebra $\cal G$ with
rank $D$. The squared length of the root vectors is normalized to two.
In this normalization, the weight lattice $\Lambda_W(\cal G)$ is
just the dual lattice of $\Lambda_R(\cal G)$. Instead of $X^I(\sigma)$,
we may use a new string coordinate $Z^I(\sigma)$ defined by
$Z^I(\sigma)=M^{IJ}X^J(\sigma)$, where $M^{IJ}$ is a constant matrix.
Then, $Z^I(\sigma)$ obeys the following boundary condition:
\begin{eqnarray}
Z^I(\sigma^1+1,\sigma^2) &=& u^{IJ} Z^J(\sigma^1,\sigma^2) +
                             \pi M^{IJ} w^J\ ,\nonumber\\
Z^I(\sigma^1,\sigma^2+1) &=&
                \widetilde{u}^{IJ} Z^J(\sigma^1,\sigma^2) +
                             \pi M^{IJ} \widetilde{w}^J\ ,
\label{zet bc}
\end{eqnarray}
where $u^{IJ}=(MUM^{-1})^{IJ}$ and
      $\widetilde{u}^{IJ}=(M \widetilde{U} M^{-1})^{IJ}$.
It should be noted that $u^{IJ}$ and $\widetilde{u}^{IJ}$
are not in general
orthogonal matrices, although $U^{IJ}$ and $\widetilde{U}^{IJ}$ are.
We may choose $M^{IJ}$ such that $M^{IJ} w^J$
(and also $M^{IJ} \widetilde{w}^J$) belongs to the lattice
$\Lambda_R(\cal G)$.
This is always possible by appropriately choosing the constant matrix
$M^{IJ}$. We will restrict our considerations to the case that
both $u^{IJ}$ and $\widetilde{u}^{IJ}$ are orthogonal matrices, otherwise
the following discussion will be invalid. We can then apply the results
of ref.\cite{toroidal} to our problem.
Let us introduce a new field $\phi(\sigma^1,\sigma^2)$ defined by
\begin{equation}
\phi(\sigma^1,\sigma^2)\ =\ \exp \left\{
         i2Z^I(\sigma^1,\sigma^2) H^I \right\}\ ,
\label{def phi}
\end{equation}
where $H^I$ is a generator of the Cartan subalgebra of $\cal G$
and is normalized such that Tr$(H^IH^J)=\delta^{IJ}$.
We note that $\phi(\sigma^1,\sigma^2)$ is a mapping from $\Sigma$
into the Cartan subgroup of the group $G$, the algebra
of which is $\cal G$.
A Wess-Zumino term \cite{WZ}-\cite{Petersen} at level one is given by
\begin{equation}
\Gamma_{WZ}[ \widetilde{\phi}]\ =\
    -\frac{i}{12\pi} \int_M Tr\left( \widetilde{\phi}^{-1}
      d\widetilde{\phi} \right)^3\ ,
\label{def WZW}
\end{equation}
where $M$ is a three dimensional manifold whose boundary is $\Sigma$
and $\phi$ is extended to a mapping $\widetilde{\phi}$ from
$M$ into $G$
with $\widetilde{\phi}\big{\vert}_\Sigma = \phi$.
The Wess-Zumino term is independent of the metric and vanishes
for any infinitesimal variation, i.e.,
\begin{equation}
\delta \Gamma_{WZ}[\widetilde{\phi}]\ =\ -\frac{i}{4\pi}\int_\Sigma
   Tr\left(\phi^{-1}\delta\phi(\phi^{-1}d\phi)^2\right)\ =\ 0\ .
\label{variation}
\end{equation}
Thus, $\Gamma_{WZ}[\widetilde{\phi}]$ will depend only on the boundary
condition (\ref{zet bc}) or (\ref{twist bc}).
We may write the Wess-Zumino term as
$\Gamma_{WZ}=\Gamma_{WZ}(U,w;\widetilde{U},\widetilde{w})$.
To this end, we will use the Polyakov-Wiegmann formula \cite{P-W},
\begin{equation}
\Gamma_{WZ}[\widetilde{\phi}_1 \widetilde{\phi}_2]\ =\
   \Gamma_{WZ}[\widetilde{\phi}_1] + \Gamma_{WZ}[\widetilde{\phi}_2]
   -\frac{i}{4\pi}\int_\Sigma
   Tr(\phi^{-1}_1d\phi_1 \,\phi_2d\phi^{-1}_2)\ .
\label{PWformula}
\end{equation}
In terms of the zero modes, the formula (\ref{PWformula})
may be written as
\begin{eqnarray}
\Gamma_{WZ}(U,w_1+w_2
    \!\!\!\!&;&\!\!\!\!\widetilde{U},\widetilde{w}_1+\widetilde{w}_2)
   = \Gamma_{WZ}(U,w_1;\widetilde{U},\widetilde{w}_1) +
     \Gamma_{WZ}(U,w_2;\widetilde{U},\widetilde{w}_2)\nonumber\\
    \!\!&-&\!\!i\pi \Bigl( w^I_1(M^{T}MU)^{IJ}\widetilde{w}^J_2
     - \widetilde{w}^I_1 \bigl(M^{T}M\widetilde{U})^{IJ}w^J_2
    \Bigr)
      \quad \mbox{mod\ $2\pi i$}.
\label{PWformula2}
\end{eqnarray}
Let us introduce an antisymmetric matrix $\Delta B^{IJ}$ and
a symmetric matrix $C^{IJ}_U$ through the relations,
\begin{equation}
 w^I \Delta B^{IJ} w'^J = w^I (M^{T}M)^{IJ} w'^J
                         \qquad \qquad \qquad \; \mbox{mod\ 2}\ ,
\label{def antisym}
\end{equation}
\begin{equation}
w^I C^{IJ}_U w'^J  =
             \frac{1}{2} w^{I} (\Delta B- U^T \Delta BU)^{IJ} w'^J
                           \qquad \mbox{mod\ 2}\ ,
\label{def sym1}
\end{equation}
for all $w^I, w'^I \in \Lambda$.
It should be noted that the antisymmetric matrix $\Delta B^{IJ}$ and
the symmetric matrix $C^{IJ}_U$ can always be defined through
the relations (\ref{def antisym}) and (\ref{def sym1}) for our choice of
the lattice $\Lambda_R(\cal G)$.
Let us write $\Gamma_{WZ}$ into the form,
\begin{eqnarray}
\Gamma_{WZ}(U,w;\widetilde{U},\widetilde{w})
   &=& i\frac{\pi}{2} w^I C^{IJ}_{\widetilde{U}}w^J
       + i\frac{\pi}{2}\widetilde{w}^I C^{IJ}_U \widetilde{w}^J
       - i\frac{\pi}{2}
          \widetilde{w}^I(U^T\Delta B\widetilde{U})^{IJ}w^J\nonumber\\
   & & - i\frac{\pi}{2}\widetilde{w}^I\Delta B^{IJ}w^J
       + \Delta \Gamma(U,w;\widetilde{U},\widetilde{w})\ .
\label{zeroWZW}
\end{eqnarray}
Then, it turns out that $\Delta \Gamma$ would be of the form
$\Delta \Gamma = -i\pi(w^I\widetilde{a}^I-\widetilde{w}^Ia^I)$
modulo $2\pi i$ for some constant vectors $a^I$
and $\widetilde{a}^I$.
It follows from eq. (\ref{shiftaction2}) that $a^I$ and $\widetilde{a}^I$
can be absorbed into the redefinition of $v^I$ and $\widetilde{v}^I$,
so that we can assume $\Delta \Gamma=0$ without loss of generality.
In the following two sections, we will see that the orbifold models
with the topological terms $S_B$ and $\Gamma_{WZ}$
correspond to those with
the antisymmetric background field
$B^{IJ} \equiv B^{IJ}_0 + \Delta B^{IJ}$,
which in general does not commute with twists $U^{IJ}$.

A simple example discussed just above is the orbifold model with
the lattice $\Lambda = \lambda \Lambda_R(\cal G)$
for some constant $\lambda$.
The matrix $M^{IJ}$ may be chosen as $M^{IJ} = \lambda^{-1} \delta^{IJ}$.
Then, the matrix $u^{IJ}$ is equal to $U^{IJ}$ and
hence is an orthogonal matrix. Thus, we can apply the above discussion
to this orbifold model.

We should make a comment on the Wess-Zumino term. The Wess-Zumino term
defined in eq.(\ref{def WZW}) might be modified
to make it well defined
\cite{F-G-K} for some orbifold models.
Our results obtained above crucially
rely on the formula (\ref{PWformula}) rather than the expression
(\ref{def WZW}) itself. Thus, our results may be valid
even if the expression (\ref{def WZW}) is ill defined. What we need is
the existence of a term which satisfies the relation (\ref{PWformula}).


\sect{ Physical Meanings of the Topological Terms}

 We have added the three 	topological terms to the kinetic term.
The total action is now given by
\footnote{
The antisymmetric background field $B_0^{IJ}$ in $S_B$
must commute with $U^{IJ}$ and $\widetilde{U}^{IJ}$, and
the last term $\Gamma_{WZ}$ cannot always be added to
the action, as noticed in the previous sections.}
\begin{equation}
S[X,g] = S_0[X,g] + S_B[X] + S_v[X] + \Gamma_{WZ}[\widetilde{\phi}]\ .
\label{totalaction}
\end{equation}
In terms of the zero modes, the last three terms
in eq.(\ref{totalaction})
can be written as
\begin{eqnarray}
S_B + S_v + \Gamma_{WZ} &=&
     i\frac{\pi}{2} w^I C^{IJ}_{\widetilde{U}}w^J
       + i\frac{\pi}{2}\widetilde{w}^I C^{IJ}_U \widetilde{w}^J
         - i\frac{\pi}{2}\widetilde{w}^I(U^TB\widetilde{U})^{IJ}w^J
          \nonumber\\
   & & - i\frac{\pi}{2}\widetilde{w}^IB^{IJ}w^J
         -i\pi\Bigl(w^I(\widetilde{U}^T)^{IJ}\widetilde{v}^J
                     - \widetilde{w}^I(U^T)^{IJ}v^J\Bigr)\ ,
\label{zeroTop}
\end{eqnarray}
where the antisymmetric matrix $B^{IJ}$ is defined by
\begin{equation}
B^{IJ} = B^{IJ}_0 + \Delta B^{IJ}.
\label{antisym}
\end{equation}
We note that the definition of $C^{IJ}_U$ in eq. (\ref{def sym1}) can
equivalently be rewritten as
\begin{equation}
w^IC^{IJ}_Uw'^J  =  \frac{1}{2}w^I(B-U^TBU)^{IJ}w'^J
                      \qquad \mbox{mod\ 2}\ ,
\label{def sym2}
\end{equation}
for all $w^I, w'^I \in \Lambda$ since $B^{IJ}_0$ commutes with $U^{IJ}$.

Since the last three terms in the action
(\ref{totalaction}) are topological
ones, they affect only on
zero mode eigenvalues. We will here clarify the effect of
the topological terms on zero mode eigenvalues from a path integral
formalism point of view. To see this, it may be instructive to recall
the Aharonov-Bohm effect in the presence of an infinitely long solenoid
\cite{AB}. If an electron moves around the solenoid,
a wave function of the
electron in general acquires a phase. In a path integral formalism
point of view, this phase is given by the classical action.
It may be natural to ask whether Aharonov-Bohm like effects occur
in our system.
Let us consider a twisted string obeying the boundary condition,
\begin{equation}
X^I(\sigma + 1, \tau) = U^{IJ}X^J(\sigma, \tau) + \pi w^I,
\label{Minkowski bc}
\end{equation}
where $\tau$ is a \lq\lq time" coordinate \footnote{
Since the topological terms are independent of the
world sheet metric, the following
arguments do not depend on the signature of the metric.
We will here take the Euclidean signature.}.
Suppose that the twisted string moves around the torus, say,
from a point \{$X^I$\} at $\tau = 0$ to a point
\{$X^I + \pi \widetilde{w}^I$\}
at $\tau = 1$ for some $\widetilde{w}^I \in \Lambda$, i.e.,
\begin{equation}
X^I(\sigma, 1) = X^I(\sigma, 0) + \pi \widetilde{w}^I.
\label{tau bc}
\end{equation}
The consistency of the boundary conditions (\ref{Minkowski bc})
and (\ref{tau bc}) requires that $\widetilde{w}^I$ must belong to
\begin{equation}
\widetilde{w}^I \in \Lambda_U \equiv \{\ w^I \in \Lambda\ |\ w^I
= U^{IJ}w^J\ \}\ .
\label{LambdaU}
\end{equation}
When the twisted string moves around the torus, the wave function
$\Psi (x^I)$ of the string can acquire a phase
$\exp\{-S_B - S_v - \Gamma_{WZ}\}$ in a similar way to the electron
moving around the solenoid. Thus we may have\footnote{We have taken
$\widetilde{U} = {\bf 1}$ and $\widetilde{v} = 0$.}
\begin{equation}
\Psi(x^I+\pi \widetilde{w}^I) = \rho\  \Psi(x^I)\ ,
\label{AB}
\end{equation}
where
\begin{equation}
\rho = \exp\{-i\pi \widetilde{w}^I(-B^{IJ}w^J + v^I + s^I_U)\}\ .
\label{rho}
\end{equation}
The constant vector $s^I_U$ with $s^I_U = U^{IJ} s^J_U$ is defined
through the relation\footnote{The existence of such a vector $s^I_U$
is guaranteed by the fact that
$\frac{1}{2}(w + w')^I C^{IJ}_U (w + w')^J
 = \frac{1}{2}w^IC^{IJ}_Uw^J + \frac{1}{2}w'^IC^{IJ}_Uw'^J$
modulo 2 for all $w^I, w'^I \in \Lambda_U$.},
\begin{equation}
\widetilde{w}^I s^I_U = \frac{1}{2}
       \widetilde{w}^I C^{IJ}_U \widetilde{w}^J\
              \qquad \mbox{mod\ 2}\ ,
\label{s-def}
\end{equation}
for $\widetilde{w}^I \in \Lambda_U$.
To see a physical implication of eq.(\ref{AB}), we note that
the left hand side of eq.(\ref{AB}) may be expressed as
\begin{equation}
\Psi(x^I+\pi \widetilde{w}^I) =
       \exp\{-i\pi \widetilde{w}^I\widehat{p}^I_{_{\!/\!/}}\}\
                    \Psi(x^I)\ ,
\label{transl}
\end{equation}
where
$\widehat{p}^I_{_{\!/\!/}}$
is a canonical momentum operator
restricted to the $U$-invariant subspace, i.e.,
{}$\widehat{p}^I_{_{\!/\!/}} = U^{IJ} \widehat{p}^J_{_{\!/\!/}}$.
Comparing eq.(\ref{AB}) with eq.(\ref{transl}), we conclude that
\begin{equation}
\widehat{p}^I_{_{\!/\! /}} \in 2{\Lambda_U}^{\ast}
 - B^{IJ}w^J + v^I + s^I_U\ ,
\label{p-eigenv}
\end{equation}
where ${\Lambda_U}^{\ast}$ is the dual lattice of $\Lambda_U$.
In the next section, we will verify the result (\ref{p-eigenv})
in the operator formalism.

To make the correspondence clear between the path integral formalism
and the operator one, we will here give an expression of
a one-loop partition function in the path integral formalism.
In the next section, we will see that the same partition function can be
obtained from the operator formalism. The one-loop partition function
$Z(\tau)$ of the orbifold model will be of the form,
\begin{equation}
Z(\tau) = \frac{1}{|P|}\sum_{{g,h \in P}\atop{gh=hg}}Z(g,h;\tau)\ ,
\label{partition}
\end{equation}
where $\tau$ is the modular parameter and $|P|$
is the order of the group $P$.
The $Z(g,h;\tau)$ denotes the contribution from the string
twisted by $g = ( U, v ) \ (h = (\widetilde{U}, \widetilde{v}) )$
in the $\sigma^1$- ($\sigma^2$-) direction.
Since $U \widetilde{U} = \widetilde{U}U$,
we can take the following coordinate system:
\begin{eqnarray}
U^{IJ} &=&
  \left(
  \begin{array}{cccc}
    \delta^{i_1j_1} & 0 & 0 & 0 \\
    0 & \delta^{i_2j_2} & 0 & 0 \\
    0 & 0 & u^{i_3j_3}_3 & 0   \\
    0 & 0 & 0 & u^{i_4j_4}_4
  \end{array}
  \right)^{IJ}\ ,\nonumber\\
\widetilde{U}^{IJ} &=&
  \left(
  \begin{array}{cccc}
    \delta^{i_1j_1} & 0 & 0 & 0 \\
    0 & \widetilde{u}^{i_2j_2}_2 & 0 & 0 \\
    0 & 0 & \delta^{i_3j_3} & 0   \\
    0 & 0 & 0 & \widetilde{u}^{i_4j_4}_4
  \end{array}
  \right)^{IJ}\ ,\nonumber\\
X^I &=& ( X^{i_1},X^{i_2},X^{i_3},X^{i_4} )\ ,
\label{UVsystem}
\end{eqnarray}
where $i_k,j_k = 1,2,\cdots,d_k (k=1,2,3,4)$ and
$I,J=1,2,\cdots,D\ (D=d_1+d_2+d_3+d_4).$
Since the topological terms contribute only to the zero mode part of
$Z(g,h;\tau)$, it will be enough to give an expression of
the zero mode part $Z(g,h;\tau)_{{\rm zero}}$ for our purpose.
The result is \footnote{For the details to derive eq.(\ref{pathint}),
see ref.\cite{I-N-T}.}
\begin{eqnarray}
 Z(g,h;\tau)_{{\rm zero}}
&=&
    \int_{\pi \Lambda} dx^I
    \sum_{w^I\in \Lambda}
    \sum_{\widetilde{w}^I \in \Lambda}
    \delta_{(1-\widetilde{U})^{IJ}w^J,(1-U)^{IJ}\widetilde{w}^J}
    \ \delta\Bigl(x^{i_2}
        - \pi \Bigl(\frac{1}{1-\widetilde{u}_2}\Bigr)
        ^{i_2j_2}\widetilde{w}^{j_2}\Bigr)\nonumber\\
& & \times\delta\Bigl(x^{i_3}
        - \pi \Bigl(\frac{1}{1-u_3}\Bigr)
      ^{i_3j_3}w^{j_3}\Bigr)
    \delta\Bigl(x^{i_4}
        - \pi \Bigl(\frac{1}{1-u_4}\Bigr)
      ^{i_4j_4}w^{j_4}\Bigr)\nonumber\\
& & \times
    \exp\biggl\{ -\frac{\pi}{2}
     \Bigl[\
       \frac{|\tau|^2}{\tau_2}(w^{i_1})^2 + 2\frac{\tau_1}{\tau_2}
       w^{i_1}\widetilde{w}^{i_1} +
       \frac{1}{\tau_2}(\widetilde{w}^{i_1})^2\
     \Bigr]  \nonumber\\
& &  \qquad -i\frac{\pi}{2}w^IC^{IJ}_{\widetilde{U}}w^J
       -i\frac{\pi}{2}\widetilde{w}^IC^{IJ}_U\widetilde{w}^J
       +i\frac{\pi}{2}\widetilde{w}^I(U^TB\widetilde{U})^{IJ}w^J
           \nonumber\\
& &  \qquad   +i\frac{\pi}{2}\widetilde{w}^IB^{IJ}w^J
               +i\pi\Bigl(w^I(\widetilde{U}^T)^{IJ}\widetilde{v}^J
                 - \widetilde{w}^I(U^T)^{IJ}v^J\Bigr) \biggr\}\ ,
\label{pathint}
\end{eqnarray}
where $\tau =\tau_1+i\tau_2$.
In the next section, we will see that the phase
appearing in eq.(\ref{pathint}), which comes from the topological terms
in the path integral formalism, has a quite different origin
in the operator formalism.


\sect{Operator Formalism}

 In this section, we shall generalize the results of ref.\cite{top}
to orbifold models with shifts and clarify physical meanings of the
topological terms discussed in the previous sections, from an operator
formalism point of view. Although all technical tools have already been
developed in ref.\cite{top}, it may worth while adding this section since
the generalization requires lengthy nontrivial calculations and since the
correspondence between the path integral formalism and the operator one is
quite nontrivial. For details and notations in this section,
see ref.\cite{top}.

We first construct the zero mode part of the Hilbert space in the
$g = (U, v)$ twisted sector. It should be emphasized that any topological
term does not contribute to the Hamiltonian as well as the equation of
motion. In the operator formalism, the antisymmetric background field
$B^{IJ}$ does not appear explicitly in the Hamiltonian but in the following
commutation relations of zero modes:
\beqs
 [\ \widehat{x}^I_L\ ,\ \widehat{x}^J_L\ ] &=&
         i\pi \Bigl(B - \sum^{N}_{m =1}\frac{m}{N}
         (U^{-m} - U^{m})\Bigr)^{IJ}\ ,\nonumber\\
 {[\ \widehat{x}^I_R\ ,\ \widehat{x}^J_R\ ]} &=&
         i\pi \Bigl(B + \sum^{N}_{m =1}\frac{m}{N}
         (U^{-m} - U^{m})\Bigr)^{IJ}\ ,\nonumber\\
 {[\ \widehat{x}^I_L\ ,\ \widehat{x}^J_R\ ]} &=&
         i\pi (1 - B)^{IJ}\ .
\label{quant1}
\eeqs
where $N$ is the smallest positive integer such that $U^N = {\bf 1}$. The
commutation relations (\ref{quant1}) are derived from the requirement of
the duality of amplitudes. All anomalous features of the orbifold models
originate in eqs.(\ref{quant1}). We should make a comment on $B^{IJ}$. The
antisymmetric background field has to satisfy
\begin{equation}
(B-U^TBU)^{IJ}w^J \in 2\Lambda^{\ast} \qquad \mbox{for all}\ \
w^I \in \Lambda\ ,
\label{consist'}
\end{equation}
which is the necessary condition for constructing
orbifold models in the operator formalism. We note that the condition
(\ref{consist'}) is satisfied for
the antisymmetric background field $B^{IJ}
= B^{IJ}_0 + \Delta B^{IJ}$ introduced in the path integral formalism.
A key to construct the Hilbert space of the zero modes is the following
operator:
\beqs
I_{k_L,k_R} &\equiv&
  \xi_{k_L,k_R} \exp\{-ik^I_LU^{IJ}\widehat{x}^J_L
    -ik^I_RU^{IJ}\widehat{x}^J_R\}
  \exp\{ik^I_L\widehat{x}^I_L + ik^I_R\widehat{x}^I_R\}   \nonumber\\
& &\quad \times
    \exp\{i2\pi({k^I_L}_{_{/\! /}}\widehat{p}^I_{L_{/\! /}}
                      - {k^I_R}_{_{/\! /}}\widehat{p}^I_{R_{/\! /}})\},
\label{identity}
\eeqs
where
\beqs
\xi_{k_L,k_R}  &\equiv&
\exp\Bigl\{i\pi \Bigl((k^I_{L_{/\! /}})^2 - (k^I_{R_{/\!/}})^2\Bigr)
\nonumber\\
 & &\quad  - i\frac{\pi}{2}(k_L-k_R)^I(UC_UU^T)^{IJ}(k_L-k_R)^J
       + i\pi (k_L-k_R)^Iv^I \Bigr\}\ .
\eeqs
The ($k^I_L, k^I_R$) is the left- and right-moving momentum and
($k^I_{L_{/\! /}}, k^I_{R_{/\! /}}$) denotes the momentum restricted to
the $U$- invariant subspace. It is not difficult to show that the operator
$I_{k_L,k_R}$ commutes with all physical operators. Thus, $I_{k_L,k_R}$
must be a $c$- number. We can further show that $I_{k_L,k_R}$ satisfies
the relation $I_{k_L,k_R}I_{k'_L,k'_R} = I_{k_L+k'_L,k_R+k'_R}$.
Hence, $I_{k_L,k_R}$ would be of the form
$I_{k_L,k_R} = \exp\{ik^I_La^I_L - ik^I_Ra^I_R\}$
for some constant vectors $a^I_L$ and $a^I_R$.
These constant vectors can be determined by requiring the relation
$gV(k_L,k_R;z)g^{\dag} = V(k_L,k_R;e^{2\pi i}z)$, where $V(k_L,k_R;z)$
is a vertex operator. Then we conclude that $a^I_L = a^I_R = 0$, i.e.,
\begin{equation}
I_{k_L,k_R} = 1\ .
\label{identity2}
\end{equation}
This identity gives constraints on the eigenvalues of the (restricted)
momentum ${\widehat{p}^I}_{_{\!/\! /}}$ and the winding number
$\widehat{w}^I$. The identity (\ref{identity2}) must be satisfied for all
momenta  ($k^I_L, k^I_R$). It is easy to show that the identity
(\ref{identity2}) for $k^I_L - k^I_R \in \Lambda_U$ reduces to
\footnote{The definition of $\widehat{p}^I_{_{\!/\! /}}$ in
eq.(\ref{identity3}) is  $\widehat{p}^I_{_{\!/\! /}}
= \widehat{p}^I_{L_{/\! /}} + \widehat{p}^I_{R_{/\! /}}$, which is
different from that in ref.\cite{top} by
$(B^{IJ}\widehat{w}^J)_{_{\!/\! /}}$.}
\begin{equation}
\exp\Bigl\{i\pi w^I_{_{\!/\! /}}(\widehat{p}^I+B^{IJ}\widehat{w}^J
            -v^I-s^I_U)_{_{\!/\! /}}+i\pi k^I\widehat{w}^I\Bigr\}
     = 1\ ,
\label{identity3}
\end{equation}
for all $w^I_{_{\!/\! /}} \in \Lambda_U$ and $k^I \in 2\Lambda^{*}$.
The identity (\ref{identity3}) implies that
\begin{eqnarray}
& & (\widehat{p}^I+B^{IJ}\widehat{w}^J-v^I-s^I_U)_{_{\!/\! /}}
\in 2{\Lambda_U}^{*},\nonumber\\
& & \qquad \qquad \widehat{w}^I \in \Lambda.
\label{eigenvs}
\end{eqnarray}
This agrees with the result obtained in the previous section. The zero mode
eigenstates can be labeled by the eigenvalues of
$({\widehat{p}^I + B^{IJ}\widehat{w}^J})_{_{\!/\! /}}$ and $\widehat{w}^I$
as $|(k^I+v^I+s^I_U)_{_{\!/\! /}},w^I>$ for
$k^I_{_{\!/\! /}} \in 2{\Lambda_U}^{\ast}$ and $w^I \in \Lambda$.
Due to the existence of the identity operator (\ref{identity2}),
all states $|(k^I+v^I+s^I_U)_{_{\!/\! /}},w^I>$ are not independent.
The inner product of the states is quite complicated and is given by
\begin{eqnarray}
& &<(k^I+v^I+s^I_U)_{_{\!/\! /}}, w^I | (k'^I+v^I+s^I_U)_{_{\!/\! /}},
w'^I >   \nonumber\\
& &\qquad = \sum_{\widetilde{w}^I \in \Lambda/\Lambda_U}
  \rho(k,w;\widetilde{w})\
  \delta_{{k'}^I_{_{\!\!\!\!/\! /}},
    (k^I+[U,B]^{IJ}\widetilde{w}^J)_{_{\!/\! /}}}
  \delta_{w'^I,w^I+(1-U)^{IJ}\widetilde{w}^J}\ ,
\label{innerpro}
\end{eqnarray}
where
\beqs
 \rho(k,w;\widetilde{w})
& = &
   \exp\biggl\{-i\frac{\pi}{2}\widetilde{w}^IC^{IJ}_U\widetilde{w}^J
   + i\pi {\widetilde{w}}^I_{_{\!/\! /}}(k^I+v^I+s^I_U)_{_{\!/\! /}}
   + i\frac{\pi}{2}\widetilde{w}^I(U^TB)^{IJ} w^J_{_{\!/\! /}}\nonumber\\
& &  \qquad - i\frac{\pi}{2}\widetilde{w}^I_{_{\!/\! /}}B^{IJ}w^J
 +
i\frac{\pi}{2}\widetilde{w}^I(B-U^TBU)^{IJ}\Bigl(\frac{1}{1-\widetilde{U}}
   \Bigr)^{JK}_{\perp}w^K_{\perp}\nonumber\\
& & \qquad - i\frac{\pi}{2}\widetilde{w}^I(B-U^TBU)^{IJ}
\widetilde{w}^J_{_{\!/\! /}} - i\pi
\widetilde{w}^I (U^T)^{IJ} v^J \biggr\}\ .
\label{rho'}
\eeqs
Here, $w^I$ has been decomposed into two subspaces, i.e.,
$w^I = (w_{_{\!/\! /}},w_{\perp})$ and $\Lambda/\Lambda_U$ in
eq.(\ref{innerpro}) denotes the set of the independent lattice points of
$\Lambda$ with the identification
$\widetilde{w}^I \sim \widetilde{w}^I + \beta^I$
for all ${\beta}^I \in \Lambda_U$.

Another nontrivial feature is an anomalous action of twisted operators on
vertex operators. Let $h = (\widetilde{U}, \widetilde{v})$ be an element of
$P$. The action of $h$ on a vertex operator is found to be
\begin{eqnarray}
h V(k_L,k_R;z) h^{\dag}
 = \eta(k_L,k_R;h)\ e^{i\pi(k_L-k_R)^I\widetilde{v}^I}
           \ V(\widetilde{U}^Tk_L,\widetilde{U}^Tk_R;z)\ ,
\label{hVh}
\end{eqnarray}
where
\beq
\eta(k_L,k_R;h) = \exp \left\{-i\frac{\pi}{2}(k_L-k_R)^I
 (\widetilde{U}C_{\widetilde{U}}\widetilde{U}^T)^{IJ} (k_L-k_R)^J\right\}\ .
\label{eta}
\eeq
The anomalous phase $\eta(k_L,k_R;h)$ is required from the consistency with
the commutation relations (\ref{quant1}). The nontrivial phases in
eqs.(\ref{rho'}) and (\ref{eta}) play a crucial role in the following
discussion.

We are now ready to compute the one-loop partition function. Let
$Z(g,h;\tau)$ be the partition function of the $g$-sector twisted by $h$
which is defined, in the operator formalism, by
\beq
Z(g,h;\tau) =
{\rm Tr}\Bigl[\:h\:e^{i2\pi \tau(L_0-D/24)-i2\pi \overline{\tau}
    (\overline{L}_0-D/24)}\:\Bigr]_{g-{\rm sector}}\ ,
\label{partitiongh}
\eeq
where $L_0$ $(\overline{L}_0)$ is the Virasoro zero mode operator
of the left- (right-) mover.
The zero mode part of $Z(g,h;\tau)$ is computed as
\footnote{The prefactor in eq.(\ref{partzero1}) has been chosen to agree
with $Z(g,h;\tau)_{{\rm zero}}$ given in the path integral formalism.}
\begin{eqnarray}
Z(g,h;\tau)_{{\rm zero}}
&=& \pi^{d_1}\det(1-\widetilde{u}_2)\ (2\tau_2)^{d_1/2}\nonumber\\
& & \quad \times
    \sum_{k^I_{_{\!/\! /}}\in 2{\Lambda_U}^{*}}
    \sum_{w^I \in \Lambda/(1-U)\Lambda}
    \sum_{\widetilde{w}^I \in \Lambda/\Lambda_U}
    {\cal A}(g,h;\tau;k_{_{\!/\! /}},w,\widetilde{w}) \ ,
\label{partzero1}
\end{eqnarray}
where
\begin{eqnarray}
\lefteqn{
{\cal A}(g,h;\tau;k_{_{\!/\! /}},w,\widetilde{w})}\nonumber\\
 & & = \delta_{(1-\widetilde{U})^{IJ}_{_{\!/\! /}}
      (k^J-B^{JK}w^K+v^J+s^J_U)_{_{\!/\! /}},0}\,
     \delta_{(1-\widetilde{U})^{IJ}w^J,(1-U)^{IJ}\widetilde{w}^J}
\nonumber\\
 & & \ \times
 \exp\biggl\{i\frac{\pi}{4}\tau
    (k^I-B^{IJ}w^J+v^I+s^I_U+w^I)^2_{_{\!/\! /}}
  -i\frac{\pi}{4}\overline{\tau}
    (k^I-B^{IJ}w^J+v^I+s^I_U-w^I)^2_{_{\!/\! /}}
    \nonumber\\
 & & \qquad  -i\frac{\pi}{2}w^IC^{IJ}_{\widetilde{U}}w^J
     -i\frac{\pi}{2}\widetilde{w}^IC^{IJ}_U\widetilde{w}^J
     + i\frac{\pi}{2}\widetilde{w}^I(U^TB\widetilde{U})^{IJ}w^J
    + i\frac{\pi}{2}\widetilde{w}^IB^{IJ}w^J  \nonumber\\
 & &  \qquad
     + i\pi\widetilde{w}^I_{_{\!/\! /}}
        (k^I-B^{IJ}w^J+v^I+s^I_U)_{_{\!/\! /}}
     +i\pi w^I(\widetilde{U}^T)^{IJ}\widetilde{v}^J
     -i\pi\widetilde{w}^I(U^T)^{IJ}v^J \biggr\}\ .
\label{partzero1'}
\end{eqnarray}
It is not difficult to verify that $Z(g,h;\tau)$ in eq.(\ref{partzero1})
is identical to the expression (\ref{pathint}) computed in the path
integral formalism. All formulas we need to prove the equivalence are given
in the appendix of ref.\cite{top}.

What we wish to stress is that the equivalence of the partition functions
is quite nontrivial because the phases in the partition functions have
quite different origins in the path integral formalism and
in the operator one: In
the path integral formalism, the phase comes from the topological terms
which have been added to the kinetic term, while in the operator formalism
the phase originates in the nontrivial phases in eqs.(\ref{innerpro})
and (\ref{hVh}).


\sect{Discussion}

 We have seen that the toroidal orbifold models have topologically quite
rich structures. We have studied the orbifold models with
the topological terms $S_B$, $S_v$ and $\Gamma_{WZ}$.
Adding $\Gamma_{WZ}$ together with $S_B$ to the kinetic term is
interpreted as the incorporation of the antisymmetric background field
$B^{IJ}=B^{IJ}_0 + \Delta B^{IJ}$ in orbifold models, where
$\Delta B^{IJ}$ denotes a topologically nontrivial part of $B^{IJ}$.
Adding $S_v$ to the kinetic term is interpreted as the incorporation of
the shift $v^I$ in orbifold models. These interpretations have been
verified by comparing the partition functions computed from both the path
integral formalism and the operator one.

If we add a new conformally invariant term to the action, we may have a
new modular invariant orbifold model, in a path integral formalism point
of view. One might add the following term to the action:
\begin{equation}
S_{A}\ =\ i\int^1_0 d^2\sigma \frac{1}{2\pi}A^{IJ}
\varepsilon^{\alpha \beta}\partial_\alpha V^I
\partial_\beta V^J \ ,
\label{A^{IJ}}
\end{equation}
where $A^{IJ}$ is an antisymmetric constant matrix satisfying the condition,
\begin{equation}
 [\ A\ ,\ U\ ]\ =\ 0   \qquad\mbox{for all $U \in P$}\ ,
\label{A-U}
\end{equation}
and $V^I$ is the external field introduced in section 2.
The term $S_A$ is conformally, more precisely, topologically invariant
and hence does not destroy modular invariance of the partition function.
Adding $S_A$ to the action produces an additional phase factor in the
partition function. We have, however, failed to find any orbifold model
which produces the same partition function in the operator formalism.
We have not known whether the orbifold model with the topological term
(\ref{A^{IJ}}) leads to a consistent model from an operator formalism
point of view, although it gives a modular invariant partition function
in the path integral formalism.

It would be of great interest to look for conformally or topologically
invariant terms in string theory compactified on more general manifolds.


\vspace{1cm}
\begin{center}
{\Large\bf Acknowledgements \par}
\end{center}

We would like to thank J. O. Madsen for collaboration
at an early stage of this work.


\end{document}